\documentclass[%
 reprint,
superscriptaddress,
 amsmath,amssymb,
 aps,
pra,
floatfix,
]{revtex4-2}

\usepackage{multirow}
\usepackage{chemfig}
\usepackage{xcolor}
\usepackage{graphicx}
\usepackage{dcolumn}
\usepackage{bm}
\usepackage{hyperref}
\hypersetup{colorlinks=true, urlcolor=blue, linkcolor=blue, citecolor=blue} 

\begin{document}

\title{Electron impact fragmentation dynamics of Carbonyl sulfide:\\A combined experimental and theoretical study}

\author{Soumya Ghosh}
\email{evan.soumya@gmail.com}
 \affiliation{Department of Physical Sciences, Indian Institute of Science Education and Research Kolkata, Mohanpur-741246, India}
 
\author{Narayan Kundu}%
\email{kundu.narayan1995@gmail.com}
\affiliation{Department of Physical Sciences, Indian Institute of Science Education and Research Kolkata, Mohanpur-741246, India}
\affiliation{University of Kassel, Institute of Physics, Heinrich-Plett-Str. 40, 34132 Kassel, Germany}%

\author{Aryya Ghosh}
\email{aryya.ghosh@ashoka.edu.in}
\affiliation{Department of Chemistry, Ashoka University, Sonipat, Haryana, 131029, India}%

\author{Dhananjay Nandi}
\email{dhananjay@iiserkol.ac.in}
\affiliation{Department of Physical Sciences, Indian Institute of Science Education and Research Kolkata, Mohanpur-741246, India}%
\affiliation{Center for Atomic, Molecular and Optical Sciences \& Technologies, Joint initiative of IIT Tirupati \& IISER Tirupati, Yerpedu, 517619, Andhra Pradesh, India}


\begin{abstract}
In this study, we examine the interactions of low- to intermediate-energy electrons ($0-45$ eV) with carbonyl sulfide (OCS). These collisions lead to the formation of several anionic fragments, including C$^-$, O$^-$, S$^-$, and SO$^-$. When the incident electron energy is below the first ionization potential of the molecule, dissociative electron attachment (DEA) process dominates, primarily yielding O$^-$ and S$^-$ fragments. At higher energies, beyond the ionization potential, ion-pair dissociation (IPD) becomes the dominant process, resulting in the emergence of additional fragments such as C$^-$ and SO$^-$. This leads to an increasingly intricate mechanism, necessitating a detailed analysis to elucidate the ion-pair dissociation pathways. The absolute cross section for S$^-$ ions has been determined using the well-established relative flow technique. Theoretical cross sections are calculated using the multi-configurational time-dependent hartree (MCTDH) method, with each potential energy curve obtained from equation-of-motion coupled-cluster singles and doubles (EOM-CCSD) calculations. The computed values are in excellent agreement with the experimental data. The analysis reveals contributions from both linear and bent anionic resonant states. Due to low count rates, only relative cross section curves have been obtained for the O$^-$ and SO$^-$ ions. At higher energies, the ion pair thresholds are evaluated using the Wannier threshold law, yielding values consistent with those derived from thermochemical data. 

\end{abstract}

\maketitle

\section{\label{sec:introduction}Introduction}

Low- and intermediate-energy electrons ($0-45$ eV) are highly effective at inducing bond cleavage in neutral molecules, producing anionic fragments through two principal pathways: Dissociative electron attachment (DEA) and Ion-pair dissociation (IPD).  In DEA, an incident electron is resonantly captured to form a short-lived transient negative ion (referred to as a negative ion resonant state); subsequent decay of this metastable species yields a neutral fragment and a negative ion.  The probability of observing a specific anion is controlled by the electron-capture cross section and by the lifetime of the transient anion prior to fragmentation. 

On the other hand, the IPD proceeds by a different mechanism: the electron deposits sufficient energy to promote the molecule to a super-excited state that dissociates into a pair of oppositely charged ions. The DEA process dominates at the lower electron energies ($0-15$ eV), whereas the IPD becomes increasingly important as the incident energy approaches the intermediate regime ($15-45$ eV).  Quantifying these channels and their energy dependence is essential for modelling electron-driven chemistry in environments ranging from technological plasmas to planetary atmospheres~\cite{Lu:2001,Christophorou:1996, Park:2007}.

Carbonyl sulfide (OCS) is found in a wide range of natural environments, including the atmosphere of Venus~\cite{HONG:1997}, volcanic emissions, the primordial Earth’s atmosphere~\cite{Ueno:2009}, stellar interiors~\cite{Charnley:1997}, and dense interstellar molecular clouds~\cite{Charnley:2004}. Upon decomposition, OCS yields carbon dioxide, a greenhouse gas contributing to global warming, as well as stratospheric sulfate aerosols that can exert a net cooling effect on the climate~\cite{Bruhl:2012}. Additionally, OCS has been shown to facilitate peptide bond formation from amino acids, highlighting its prebiotic chemical significance~\cite{Luke:2004}.

Structurally, OCS is a linear triatomic molecule centered on a carbon atom, bearing electronic similarities to carbon dioxide ($\mathrm{CO_2}$) and carbon disulfide ($\mathrm{CS_2}$). Unlike $\mathrm{CO_2}$ and $\mathrm{CS_2}$, however, OCS possesses a permanent dipole moment. This key distinction has motivated extensive comparative studies among these molecules. Dillard and Franklin~\cite{Dillard:1968} first explored electron interactions with OCS and $\mathrm{CS_2}$, observing the formation of various anionic species within the $1-3.5$~eV electron energy range. Subsequent investigations by MacNeil and Thynne~\cite{Macneil:1969} further characterized the formation of negative ions in these systems. Ziesel \textit{et al.}~\cite{ziesel:1975} focused on the generation of $\mathrm{S^-}$ ions via DEA, reporting ion yield curves and cross sections over the $0.8-4.4$~eV range. Hubin-Frańskin \textit{et al.}~\cite{hubin:1976} extended this work by studying both $\mathrm{O^-}$ and $\mathrm{S^-}$ ion intensities across broader energy intervals. Additional insights into the production of $\mathrm{S^-}$ ions were provided by Abouaf and Fiquet-Fayard~\cite{Abouaf:1976}, while Tronc \textit{et al.}~\cite{Tronc:1982} investigated both ion yield curves and kinetic energy distributions of $\mathrm{S^-}$ at resonant energies. Iga and Srivastava~\cite{IGA:1995} employed a Time-of-Flight (TOF) mass spectrometer to detect $\mathrm{S^-}$, $\mathrm{O^-}$, and $\mathrm{C^-}$ ions formed via DEA to OCS, and they reported the corresponding appearance and peak energies. In a follow-up study, Iga \textit{et al.}~\cite{IGA:1996} quantified cross sections for $\mathrm{S^-}$ production over the $0-20$~eV energy range using TOF techniques. Hoffmann \textit{et al.}~\cite{Hoffmann:2008} analyzed vibrational excitation of OCS up to $8$~eV. More recently, Li \textit{et al.}~\cite{LI:2016} utilized Velocity Map Imaging (VMI) technique to identify three distinct kinetic energy bands for $\mathrm{S^-}$ ions from OCS, along with isotropic angular distributions. Kundu and Nandi~\cite{Kundu:2024} have further advanced this field by confirming the presence of Renner–Teller vibronic splitting and dipole-forbidden vibronic intensity borrowing in DEA to OCS incorporating conical time-sliced VMI study of $\mathrm{S^-}$/OCS fragmentation. A recent study by Kundu~\textit{et al.}~\cite{kundu_arxiv:2025} presented a combined experimental and theoretical investigation of DEA dynamics in carbonyl sulfide, employing velocity map imaging and equation-of-motion coupled cluster methods. Their analysis identifies a mechanism in which initially formed molecular shape resonances evolve via nonadiabatic resonant tunneling into vibronic Feshbach resonances, leading to mode-specific fragmentation pathways.

In the present work we examine the formation of S$^{-}$ fragments produced by both DEA and IPD processes in carbonyl sulphide.  Within the DEA energy window we observe O$^{-}$ and S$^{-}$ ions, with S$^{-}$ dominating the yield; at higher electron energies additional anionic products, namely, C$^{-}$ and SO$^{-}$, are detected. Absolute cross sections for S$^{-}$ are obtained with the well established relative flow technique (RFT)~\cite{Krishnakumar:1988}; absolute cross sections for the other ionic channels could not be determined because their low count rate. To complement experimental observations, energy dependent cross sections have been computed with the multi-configurational time-dependent Hartree (MCTDH) method, with each potential energy curve obtained from EOM-CCSD calculations~\cite{Krylov:2008, Bartlett:2012}. These calculations not only reproduce the absolute DEA cross sections but also identify the transient resonant states responsible for S$^{-}$ production.  Robust theoretical modelling is essential for interpreting experimental spectra and for predicting cross sections in systems or energy regimes that are difficult to probe experimentally.  The following sections describe the experimental set-up, RFT analysis, and computational methodology in detail.

\section{\label{sec:measurement}Measurement procedure}

\subsection{\label{subsec:experimental}Experimental}

The specifics of the instrument and measurement procedure are addressed in a separate publication~\cite{Chakraborty:2018}, so here it is discussed in brief. A magnetically collimated pulsed electron beam is generated from a custom-built electron gun, where a tungsten filament is heated by passing current. The emitted electrons pass through a set of electrodes maintained at specific potentials to converge the electron beam. A negative potential is applied to one electrode relative to the previous one to stop the electron flow, and then a positive pulse is applied to release the electrons producing pulse electron beam. The electron gun is placed within a pair of helmholtz coils, generating a uniform magnetic field in line with the electron beam's trajectory, collimating the electrons once they leave the electron gun. The electron gun operates at a repetition rate of $10$ kHz with a time width of $200$ ns, positioned perpendicular to the molecular beam. The effusive molecular beam is generated from a $1$ mm needle placed inside the repeller plate and the needle is electrically isolated. The spectrometer is aligned with the molecular beam axis and is perpendicular to the electron beam axis. The electrons and molecule interacts within the repeller and attractor plate forming negative ions in case of DEA and both positive and negative ions in case of IPD. The negative ions formed due to this interaction are guided to the Micro-Channel Plate (MCP) based detector using a combination of segmented and linear time-of-flight spectrometer. The spectrometer consist of a repeller plate, an attractor plate, three lens electrodes kept in einzel lens configuration, and a long field free tube. The attractor plate is designed using wire meshes with a $90\%$ ion transmission efficiency, ensuring minimal electric field penetration while allowing the majority of ions to pass. Additionally, another wire mesh is placed at the end of the long tube to prevent field penetration from the detector side.

The MCPs are arranged in a chevron (V-like) configuration to prevent ion feedback effects. Behind the MCPs, a faraday cup is positioned to collect the charge cloud and maintained at a higher potential. This faraday cup is linked to a capacitor decoupler circuit to capture the voltage fluctuations; when a charge cloud strikes the faraday cup, voltage fluctuations will occur producing signal. Following decoupling, the signal passes through an amplifier circuit for amplification. Subsequently, a Constant Fraction Discriminator (CFD) converts these fluctuations into square pulses. The CFD output serves as a stop signal for a nuclear instrumentation module (NIM)-based time-to-amplitude converter (TAC). The start pulse, synchronized with the electron gun pulse, is generated by the master pulse generator. The time difference between the start and stop pulses determines the Time-of-Flight (ToF) of the detected negative ions. The output from the TAC is fed into a Multichannel Analyzer (MCA), which interfaces with a computer through the MAESTRO software to record the Time-of-Flight spectra. The Time-of-Flight (ToF) of the ions is expressed by $\mbox{T}\propto\sqrt{m/q}$, where $m$ is the mass and $q$ is the charge of the ions. A specific mass-to-charge ratio ($m/q$) is selected and scanned across the entire electron energy range to measure the yield of the corresponding ions.\par
The Relative Flow Technique (RFT) is used to measure the absolute cross section of the fragments~\cite{Krishnakumar:1988,IGA:1996,Paul:2023,Chakraborty:2020,Orient:1983,Srivastava:1975}. RFT is a calibration procedure, where well known cross section for one anion is used to calculate the absolute cross section of other anion by comparing their intensities by keeping the experimental condition unchanged. Here, we used well established $\mbox{O}^{-}$ from $\mbox{O}_{2}$ absolute DEA cross section data~\cite{Rapp:1965} as reference. The equation used to calculate the absolute cross section for the formation of S$^{-}$ from OCS is given by

\begin{eqnarray}
\label{eqn:abs_cs}
\frac{\sigma\left(\mbox{S}^{-}/\mbox{OCS}\right)}{\sigma\left(\mbox{O}^{-}/\mbox{O}_{2}\right)}=&&\frac{\mbox{N}\left(\mbox{S}^{-}/\mbox{OCS}\right)}{\mbox{N}\left(\mbox{O}^{-}/\text{O}_{2}\right)}\times\frac{\mbox{I}_{e}\left(\mbox{O}_{2}\right)}{\mbox{I}_{e}\left(\mbox{OCS}\right)}\times\sqrt{\frac{\mbox{M}_{\text{O}_{2}}}{\mbox{M}_{\text{OCS}}}}\nonumber\\
&&\times\frac{\mbox{F}_{\text{O}_{2}}}{\mbox{F}_{\text{OCS}}}\times\frac{\mbox{K}\left(\mbox{O}^{-}/\text{O}_{2}\right)}{\mbox{K}\left(\mbox{S}^{-}/\text{OCS}\right)}
\end{eqnarray}

$\mbox{N}$ is the number of fragment ions collected for a specific time, $\mbox{F}$ is the flow rate of the gases, $\mbox{I}_{e}$ is the time averaged electron-beam current, $\mbox{M}$ is the molecular weight of the parent molecule, $\mbox{K}$ is the detection efficiency, and $\sigma$ is the absolute cross section. The overall detection efficiency, K , can be expressed as the product of three components: $\text{K} = \text{K}_1 \cdot \text{K}_2 \cdot \text{K}_3$. Here, K$_1$ represents the efficiency of extracting fragment ions from the interaction region into the spectrometer; K$_2$ is the transmission efficiency of the spectrometer in directing ions to the detector without losses; and K$_3$ is the detection efficiency of the Micro-Channel Plates (MCPs). These factors remain unchanged if the spectrometer voltages are kept optimum for total ion collection, regardless of the fragment's kinetic energy (till 4 eV) and angular distribution. Therefore, the ratio of detection efficiencies for O$^{-}$ from O$_2$ and S$^{-}$ from OCS is assumed to be unity.

This experiment is performed under ultra high vacuum condition and using highly pure commercially available carbonyl sulphide gas. The initial base pressure was $10^{-9}$ mbar before the experiment, and during the experiment, it was maintained at $10^{-7}$ mbar. The spectrometer is capable to separate out $\mbox{O}^{-}$ and $\mbox{OH}^{-}$ with mass $16$ and $17$ amu, so, there is a very low chance of impurity.

\subsection{\label{subsec:theoretical}Computational}

To support the experimental findings, we computed the absolute cross sections using the multi-configurational time-dependent hartree (MCTDH) approach. For this purpose, potential energy curves (PECs) are calculated for both the ground state and the negative ion resonant~(NIR) states up to 10~eV. These PECs are generated for two molecular configurations: a linear geometry and a bent geometry with the $\angle$OCS fixed at 135.4$^\circ$.

In the linear configuration, the C--S bond length is varied from 1.3~\AA{} to 10~\AA{}, while the C--O bond length is held fixed at 1.15~\AA{}. For the bent configuration, the C--O bond is fixed at 1.21~\AA{} during the stretching of the C--S bond over the same range.

The PECs for the ground state are computed using the highly correlated coupled-cluster singles and doubles (CCSD) method. For the electron-attached states, the equation-of-motion coupled-cluster singles and doubles (EOM-CCSD) method is employed. All calculations utilised the aug-cc-pVTZ basis set for every atom involved. Our analysis reveals that the ground state PEC of OCS in the bent geometry is approximately 3.2~eV bound.

We employ the ground-state PEC obtained from the bent geometry to construct the initial wave packet, $|\Psi_i\rangle$, which corresponds to the vibrational ground state of the electronic ground state. This initial wave packet is localised near the Franck–Condon region.

The wave packet is then vertically projected onto the electron-attached states for time propagation, and the initial condition is set as:

$$\left|\Psi(t=0)\right\rangle = |\Psi_i\rangle$$

In our calculations, the grid is defined from 1.3~\AA{} to 6~\AA{}, consisting of a total of 800 grid points. The wave packet is propagated for a total duration of 1000~fs. To prevent artificial reflection of the wave packet from the boundaries of the grid, a cubic complex absorbing potential is employed at the edges.

Non-adiabatic coupling effects are included wherever conical intersections are identified among the electron-attached states. From our computed potential energy curves (PECs), conical intersections are observed around 6 eV when using the linear configuration for generating the PECs of the electron-attached states. Additionally, for the bent configuration, conical intersections are observed near 4.8~eV and 6.8~eV.

To investigate the nuclear dynamics on the electron-attached potential energy surfaces and to compute the absolute cross section for S$^{-}$ ion formation, we numerically solved the time-dependent Schrödinger equation for the vibrational wave packets $|\Psi(t)\rangle$ propagating on the electron-attached states.

\begin{equation*}
i \left| \dot{\Psi} \right\rangle = \left[\hat{\text{T}}_{\text{N}}+\text{V}(\mathbf{r})+
\begin{pmatrix}
0 & \lambda \\
\lambda & 0
\end{pmatrix}\right]\left| \Psi \right\rangle
\end{equation*}

The operator $\hat{\text{T}}_\text{N}$ represents the nuclear kinetic energy, while $\text{V}(\mathbf{r})$ denotes the potential energy curve (PEC) of the electron-attached states. The term $\lambda$ corresponds to the non-adiabatic coupling matrix element. The absolute cross section for the S$^{-}$ ion is computed using a quantum flux formalism incorporating a complex absorbing potential (CAP), within the energy range of $0$ to $10$~eV. 

In our cross-section calculations, we employed a dipole-corrected norm for the initial wave packet $\left| \Psi_i \right\rangle$. The computed absolute cross section for S$^{-}$ ion formation, obtained using PECs for the electron-attached states in the MCTDH framework (considering both linear and bent configurations) is presented in Fig.~\ref{fig:s_cs_calc}. From the calculated cross section, distinct resonances are observed near $1$~eV, $5$~eV, $6.8$~eV, and $10$~eV.

\section{\label{sec:results}Result and discussion}

\subsection{\label{subsec:mass_spectrum}Mass spectrum}

\begin{figure*}
\centering
\includegraphics[scale=0.6]{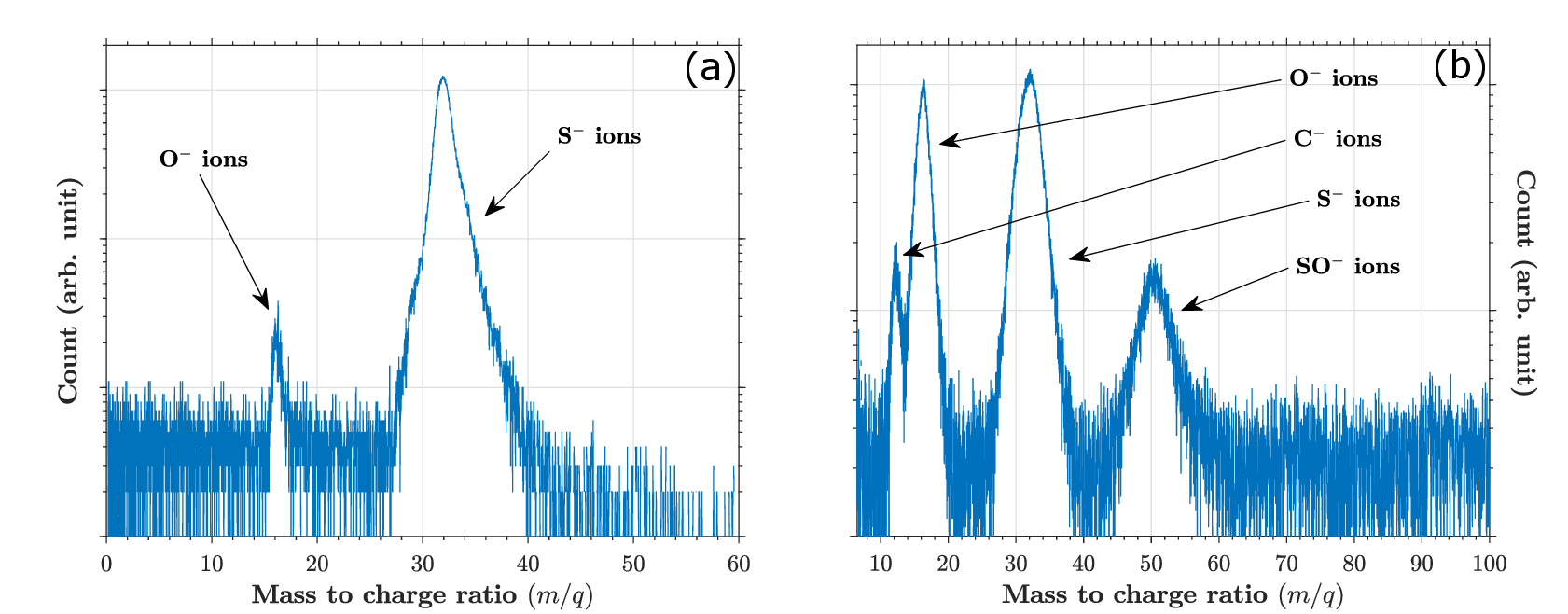}
\caption{\label{fig:ocs_masses}(a)~The mass spectra of anionic fragments generated at an incident electron energy of $11.2$ eV reveal the presence of O$^{-}$ and S$^{-}$ ions. At this energy, the O$^{-}$ ion signal is notably weak, resulting in a poor signal-to-noise ratio. (b)~The mass spectra of anionic fragments generated at an incident electron energy of $40.8$ eV reveal the presence of C$^{-}$, O$^{-}$, S$^{-}$ and SO$^{-}$ ions.}
\end{figure*}

Earlier studies by Dillard and Franklin~\cite{Dillard:1968}, MacNeil and Thynne~\cite{Macneil:1969}, Ziesel \textit{et al.}~\cite{ziesel:1975}, Hubin-Frańskin \textit{et al.}~\cite{hubin:1976}, Abouaf and Fiquet-Fayard~\cite{Abouaf:1976}, and Tronc \textit{et al.}~\cite{Tronc:1982} reported the formation of S$^-$ and O$^-$ anions resulting from low-energy electron collisions with carbonyl sulfide (OCS). Iga \textit{et al.}~\cite{IGA:1995,IGA:1996} further identified the presence of C$^-$ ions in the same energy regime. However, no evidence for the formation of other anionic species was found at low energies.

The mass spectra of electron-OCS collisions at two representative energies are shown in Fig.~\ref{fig:ocs_masses}. At 11.2~eV (Fig.~\ref{fig:ocs_masses}(a)), only S$^-$ and O$^-$ ions are detected. The O$^-$ signal exhibits a low signal-to-noise ratio, preventing reliable cross-section determination, whereas the absolute cross section is evaluated for S$^-$ ions. No trace of C$^-$, CS$^-$, CO$^-$, SO$^-$ or OCS$^-$ ions are observed at this energy. At a higher energy of 40.8~eV, well within the ion-pair formation regime (Fig.~\ref{fig:ocs_masses}(b)), additional fragments, C$^-$ and SO$^-$ are clearly identified. This is consistent with earlier findings by MacNeil and Thynne~\cite{Macneil:1969}, who reported C$^-$ ion production around 50~eV. The production of several anions is discussed below.

\subsection{\label{subsec:s_dissociation_channel}S$^{-}$ production}

\subsubsection{\label{subsubsec:s_dea_dissociation}The DEA process}

\begin{figure}[b]
\centering
\includegraphics[scale=0.58]{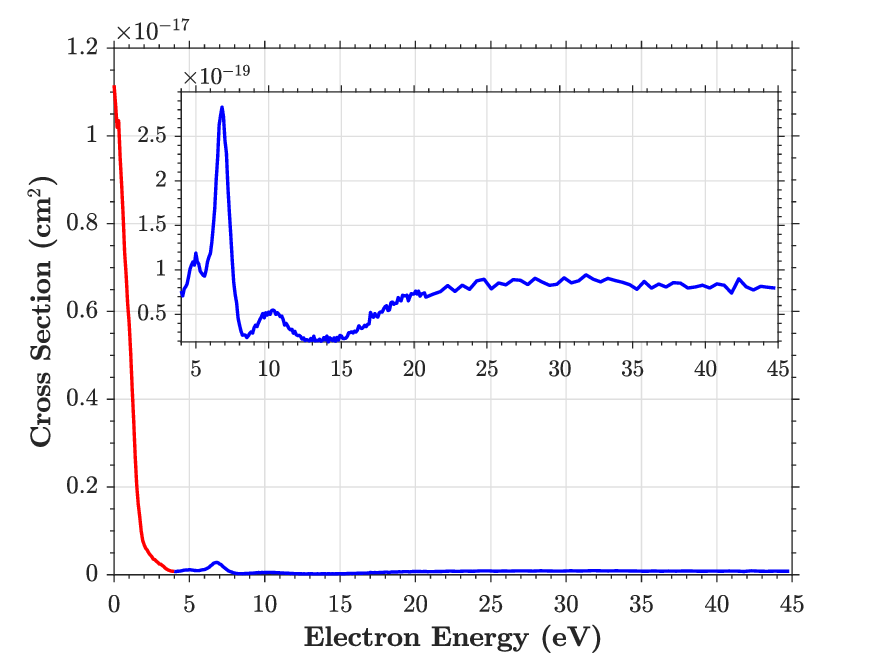}
\caption{\label{fig:s_cs_meas}The absolute cross section of the S$^{-}$ ions, corrected for electron beam intensity, is shown. Below $4$ eV (indicated in red), the electron flux is too low (comparable to the dark noise level) rendering the calculated cross sections in this range unreliable. Although a linear correction for electron beam intensity was applied, the anion production in this low-energy region does not follow a linear trend. The inset displays the absolute cross section of S$^{-}$ ions for incident electron energies between $4$ and $45$ eV, where the electron flux is sufficiently high to ensure reliable measurements.}
\end{figure}

\begin{figure*}
\centering
\includegraphics[scale=0.6]{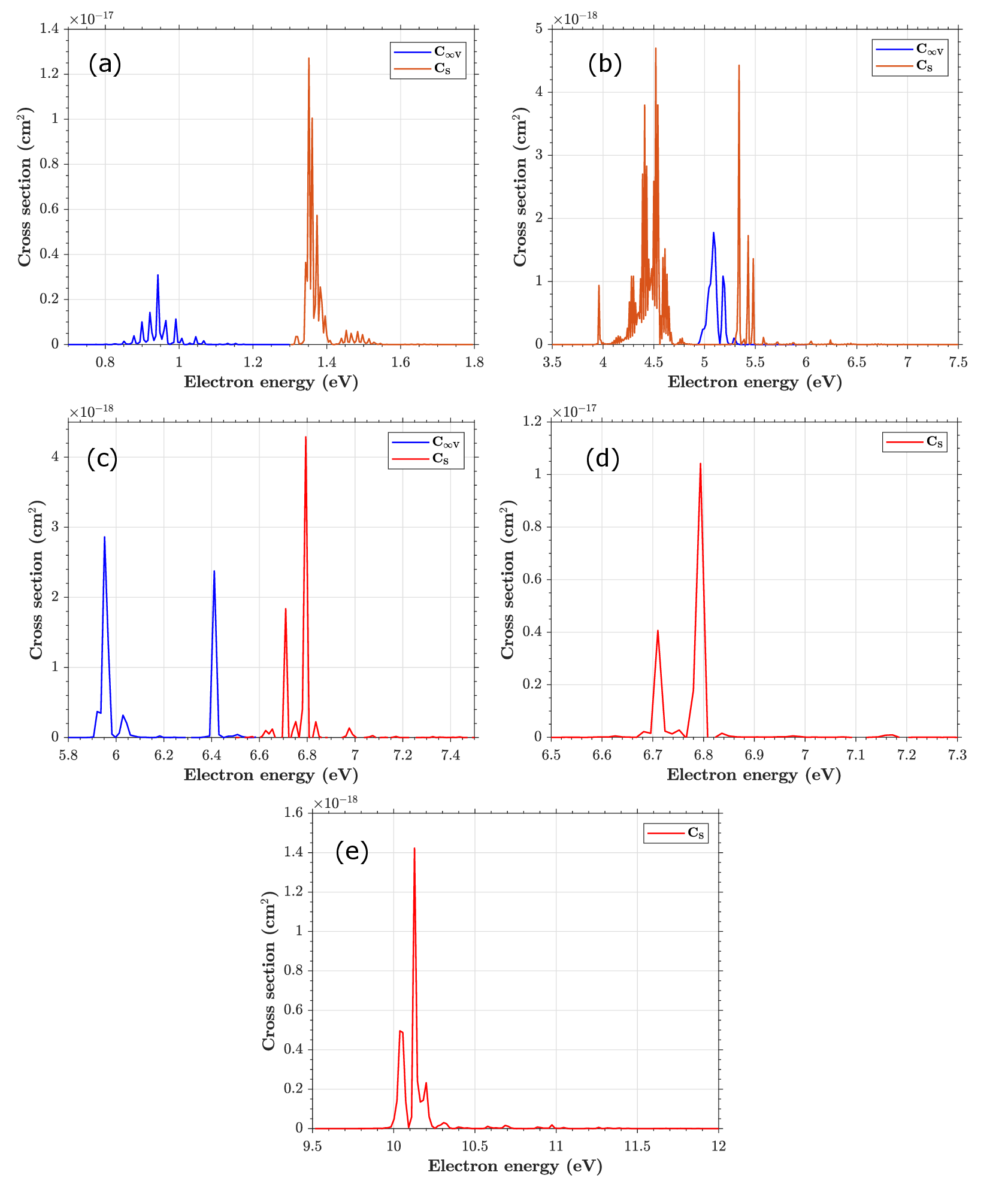}
\caption{\label{fig:s_cs_calc}The calculated cross sections for the formation of S$^{-}$ ions using the MCTDH method are shown for both C$_{\infty\text{V}}$ (linear geometry, blue curve) and C$_{\text{S}}$ ($135.4^{\circ}$ bent geometry, red curve).  
(a)~Cross section calculation near the 1.2~eV DEA resonance for both linear and bent geometries.  
(b)~Cross section calculation near the 5~eV DEA resonance for both geometries.  
(c)~Cross sections for the 6.8~eV resonance in linear and bent configurations.  
(d)~Another 6.8~eV resonance in the bent geometry is displayed separately for clarity, as it shares the same electron energy range as (c).  
(e)~Cross section near 10.2~eV shown only for the bent geometry, as no linear resonance is present at this energy.}
\end{figure*}

Several early studies~\cite{ziesel:1975,hubin:1976,Abouaf:1976,Tronc:1982,IGA:1995,IGA:1996} consistently reported a prominent low-energy resonance near 1.2~eV in electron-OCS collisions, along with the identification of multiple relatively higher-energy resonant states. In our measurements, we observed distinct resonances at $5.0$, $6.8$, and $10.2$~eV. The peak observed near zero eV is not considered reliable as a result of the very low electron beam intensity, which approaches the dark count level. Although electrons can still reach the interaction region, they lack sufficient energy to be detected at the Faraday cup. This explains the elevated ion counts despite the low electron beam intensity. While in cross section calculation, a linear correction to the electron beam intensity is applied (as described in Eqn.~\ref{eqn:abs_cs}), the production of S$^{-}$ ions in this low-energy region does not scale linearly with electron beam intensity. Consequently, rendering the cross sections below 4~eV unreliable. Fig.~\ref{fig:s_cs_meas} presents the calculated values of absolute cross section of S$^{-}$ ions using RFT. The red portion of the curve indicates the unreliable region, while the blue portion corresponds to the reliable cross section data. The reliable portion of the cross section data is presented in Table~\ref{tab2:cross_sections}. To interpret the observed resonances and understand the underlying dynamics, we employed advanced theoretical tools, the details of which is discussed in Section~\ref{subsec:theoretical}.

\begin{table*}
\caption{\label{tab2:cross_sections}Calculated cross sectional values using the relative flow technique}
\begin{ruledtabular}
\begin{tabular}{cccccccc}
Energy & $\sigma_{\text{expt}}$ & Energy & $\sigma_{\text{expt}}$ & Energy & $\sigma_{\text{expt}}$ & Energy & $\sigma_{\text{expt}}$\\
(eV) & $\left(\times 10^{-20}\text{cm}^{2}\right)$ & (eV) & $\left(\times 10^{-20}\text{cm}^{2}\right)$ & (eV) & $\left(\times 10^{-20}\text{cm}^{2}\right)$ & (eV) & $\left(\times 10^{-20}\text{cm}^{2}\right)$\\
\hline
$4.0$ & $7.62$ & $9.5$ & $4.72$ & $15.0$ & $2.57$ & $20.5$ & $7.30$ \\
$4.1$ & $7.08$ & $9.6$ & $5.10$ & $15.1$ & $2.29$ & $20.6$ & $7.33$ \\
$4.2$ & $7.87$ & $9.7$ & $4.62$ & $15.2$ & $2.29$ & $20.7$ & $7.43$ \\
$4.3$ & $8.10$ & $9.8$ & $5.19$ & $15.3$ & $2.33$ & $20.8$ & $6.94$ \\
$4.4$ & $8.43$ & $9.9$ & $4.92$ & $15.4$ & $2.52$ & $21.3$ & $7.28$ \\
$4.5$ & $9.22$ & $10.0$ & $5.27$ & $15.5$ & $2.94$ & $21.8$ & $7.56$ \\
$4.6$ & $10.09$ & $10.1$ & $5.00$ & $15.6$ & $2.96$ & $22.3$ & $8.24$ \\
$4.7$ & $10.56$ & $10.2$ & $5.43$ & $15.7$ & $3.06$ & $22.8$ & $7.59$ \\
$4.8$ & $10.90$ & $10.3$ & $5.50$ & $15.8$ & $2.91$ & $23.3$ & $8.28$ \\
$4.9$ & $10.53$ & $10.4$ & $5.40$ & $15.9$ & $3.11$ & $23.8$ & $7.82$ \\
$5.0$ & $11.89$ & $10.5$ & $4.99$ & $16.0$ & $3.02$ & $24.3$ & $8.77$ \\
$5.1$ & $10.93$ & $10.6$ & $5.20$ & $16.1$ & $3.26$ & $24.8$ & $8.95$ \\
$5.2$ & $10.66$ & $10.7$ & $5.01$ & $16.2$ & $3.72$ & $25.3$ & $7.87$ \\
$5.3$ & $9.87$ & $10.8$ & $4.75$ & $16.3$ & $3.50$ & $25.8$ & $8.53$ \\
$5.4$ & $9.63$ & $10.9$ & $4.81$ & $16.4$ & $3.41$ & $26.3$ & $8.33$ \\
$5.5$ & $9.40$ & $11.0$ & $4.42$ & $16.5$ & $3.56$ & $26.8$ & $8.91$ \\
$5.6$ & $9.30$ & $11.1$ & $3.80$ & $16.6$ & $3.78$ & $27.3$ & $8.83$ \\
$5.7$ & $9.92$ & $11.2$ & $4.00$ & $16.7$ & $3.62$ & $27.8$ & $8.35$ \\
$5.8$ & $10.90$ & $11.3$ & $3.73$ & $16.8$ & $3.88$ & $28.3$ & $9.06$ \\
$5.9$ & $11.45$ & $11.4$ & $3.42$ & $16.9$ & $3.93$ & $28.8$ & $8.63$ \\
$6.0$ & $11.85$ & $11.5$ & $3.27$ & $17.0$ & $5.22$ & $29.3$ & $8.27$ \\
$6.1$ & $13.13$ & $11.6$ & $3.33$ & $17.1$ & $4.85$ & $29.8$ & $8.39$ \\
$6.2$ & $14.93$ & $11.7$ & $2.89$ & $17.2$ & $4.68$ & $30.3$ & $9.11$ \\
$6.3$ & $16.94$ & $11.8$ & $3.10$ & $17.3$ & $5.09$ & $30.8$ & $8.55$ \\
$6.4$ & $20.17$ & $11.9$ & $2.72$ & $17.4$ & $4.82$ & $31.3$ & $8.78$ \\
$6.5$ & $22.65$ & $12.0$ & $2.62$ & $17.5$ & $4.70$ & $31.8$ & $9.45$ \\
$6.6$ & $26.27$ & $12.1$ & $2.65$ & $17.6$ & $5.18$ & $32.3$ & $8.96$ \\
$6.7$ & $27.37$ & $12.2$ & $2.51$ & $17.7$ & $5.29$ & $32.8$ & $8.66$ \\
$6.8$ & $28.29$ & $12.3$ & $2.22$ & $17.8$ & $5.25$ & $33.3$ & $9.05$ \\
$6.9$ & $27.23$ & $12.4$ & $2.36$ & $17.9$ & $5.61$ & $33.8$ & $8.81$ \\
$7.0$ & $24.56$ & $12.5$ & $2.06$ & $18.0$ & $5.86$ & $34.3$ & $8.61$ \\
$7.1$ & $23.16$ & $12.6$ & $2.17$ & $18.1$ & $5.98$ & $34.8$ & $8.34$ \\
$7.2$ & $19.40$ & $12.7$ & $2.10$ & $18.2$ & $5.44$ & $35.3$ & $7.82$ \\
$7.3$ & $16.54$ & $12.8$ & $2.03$ & $18.3$ & $5.50$ & $35.8$ & $8.71$ \\
$7.4$ & $13.90$ & $12.9$ & $2.29$ & $18.4$ & $6.31$ & $36.3$ & $7.95$ \\
$7.5$ & $11.12$ & $13.0$ & $2.07$ & $18.5$ & $6.39$ & $36.8$ & $8.41$ \\
$7.6$ & $8.63$ & $13.1$ & $2.55$ & $18.6$ & $6.14$ & $37.3$ & $8.08$ \\
$7.7$ & $7.26$ & $13.2$ & $2.02$ & $18.7$ & $6.27$ & $37.8$ & $8.57$ \\
$7.8$ & $6.31$ & $13.3$ & $2.09$ & $18.8$ & $6.26$ & $38.3$ & $8.51$ \\
$7.9$ & $4.60$ & $13.4$ & $2.04$ & $18.9$ & $6.89$ & $38.8$ & $7.97$ \\
$8.0$ & $3.82$ & $13.5$ & $2.21$ & $19.0$ & $6.62$ & $39.3$ & $8.08$ \\
$8.1$ & $3.14$ & $13.6$ & $1.99$ & $19.1$ & $6.53$ & $39.8$ & $8.28$ \\
$8.2$ & $2.67$ & $13.7$ & $2.10$ & $19.2$ & $7.15$ & $40.3$ & $7.97$ \\
$8.3$ & $2.70$ & $13.8$ & $2.40$ & $19.3$ & $7.11$ & $40.8$ & $8.43$ \\
$8.4$ & $2.68$ & $13.9$ & $2.15$ & $19.4$ & $7.08$ & $41.3$ & $8.27$ \\
$8.5$ & $2.40$ & $14.0$ & $2.56$ & $19.5$ & $7.16$ & $41.8$ & $7.40$ \\
$8.6$ & $2.58$ & $14.1$ & $1.91$ & $19.6$ & $6.53$ & $42.3$ & $9.00$ \\
$8.7$ & $2.87$ & $14.2$ & $2.38$ & $19.7$ & $6.98$ & $42.8$ & $8.11$ \\
$8.8$ & $3.04$ & $14.3$ & $2.23$ & $19.8$ & $7.34$ & $43.3$ & $7.75$ \\
$8.9$ & $2.90$ & $14.4$ & $2.06$ & $19.9$ & $7.27$ & $43.8$ & $8.16$ \\
$9.0$ & $3.50$ & $14.5$ & $2.38$ & $20.0$ & $7.31$ & $44.3$ & $8.05$ \\
$9.1$ & $3.80$ & $14.6$ & $1.99$ & $20.1$ & $7.62$ & $44.8$ & $7.96$ \\
$9.2$ & $3.62$ & $14.7$ & $2.41$ & $20.2$ & $7.29$ \\
$9.3$ & $4.06$ & $14.8$ & $2.12$ & $20.3$ & $7.61$ \\
$9.4$ & $4.41$ & $14.9$ & $2.64$ & $20.4$ & $7.05$ \\

\end{tabular}
\end{ruledtabular}
\end{table*}

\begin{table*}
\caption{\label{tab:area_under_the_curve}Area under the curve for both measured and computational values using the trapezoidal method}
\begin{ruledtabular}
\begin{tabular}{cc|ccc}
\multicolumn{2}{@{}c|@{}}{Experimental}& \multicolumn{3}{@{}c@{}}{Computational}\\
\hline
Electron  &Integrated                                    &Peak       &Molecular&Integrated\\
energy    &cross section                                 &energy     &point    &cross section\\
range (eV)&$(\times 10^{-20}\text{eV}\cdot\text{cm}^{2})$&(eV)       &group    &$(\times 10^{-20}\text{eV}\cdot\text{cm}^{2})$\\ 
\hline
\multirow{2}{*}{$-$}        & \multirow{2}{*}{$-$}       & $0.94$ & C$_{\infty\text{V}}$ & $7.24$\\
                            &                            & $1.35$ & C$_{\text{S}}$ & $23.75$\\
\hline
\multirow{3}{*}{$4.0-5.6$}  & \multirow{3}{*}{$15.36$}  & $5.09$ & C$_{\infty\text{V}}$ & $18.22$\\
                            &                            & $4.52$ & C$_{\text{S}}$ & $45.43$\\
                            &                            & $5.34$ & C$_{\text{S}}$ & $11.02$\\

\hline
\multirow{4}{*}{$5.6-8.5$}  & \multirow{4}{*}{$39.74$}  & $5.95$ & C$_{\infty\text{V}}$ & $8.76$\\
                            &                            & $6.41$ & C$_{\infty\text{V}}$ & $5.05$\\
                            &                            & $6.79$ & C$_{\text{S}}$ & $10.53$\\
                            &                            & $6.79$ & C$_{\text{S}}$ & $25.41$\\

\hline
$8.5-13$                    & $16.66$                   & $10.13$ & C$_{\text{S}}$ & $7.05$\\

\end{tabular}
\end{ruledtabular}
\end{table*}

Using the argument provided by Kundu \textit{et. al.}~\cite{kundu_arxiv:2025}, the DEA dissociation in OCS could go through a linear dissociation channel (through the linear geometry of the NIR state) or it could go through a bent dissociation channel (through $135.4^{\circ}$ bent geometry which is a stable geometry of the NIR state). The potential energy curves are presented in Ref.~\cite{kundu_arxiv:2025}, the number of dissociative states originating from the linear geometry is significantly smaller than the number of those accessible through bent molecular configurations. For linear geometries, a conical intersection is observed near 6~eV, while for bent configurations, intersections appeared around 4.8~eV and 6.8~eV. In our dynamics calculation, we have included the nonadiabatic coupling effect wherever a conical intersection is present among the NIR states. The absolute cross sections for the formation of negative ions from the resonant states were computed using the Multi-Configurational Time-Dependent Hartree (MCTDH) approach. The resulting cross sections are presented in Fig.~\ref{fig:s_cs_calc}, where the S$^{-}$ ion cross sections for both linear and bent geometries are shown. The linear geometry is depicted by the blue curve, while the bent geometry is represented by the red curve.

The $1.2$~eV resonance has been extensively studied by several authors. Ziesel~\textit{et al.}~\cite{ziesel:1975} and Abouaf~\textit{et al.}~\cite{Abouaf:1976} reported ion yield curves showing a prominent peak with a shoulder structure near this energy. Abouaf~\textit{et al.} attributed the shoulder to the vibrational excitation of neutral CO into its $\nu=1$ state and also identified higher vibrational levels. Tronc~\textit{et al.}~\cite{Tronc:1982} confirmed these findings, providing relative cross sections for the $\nu=0$ to $\nu=4$ vibrational states near threshold, and measured the kinetic energy of S$^{-}$ ions around the 1.2~eV shape resonance using a time-of-flight technique. Iga~\textit{et al.}~\cite{IGA:1996} reported an absolute cross section for S$^{-}$ production peaking at $2.6 \times 10^{-17}~\text{cm}^{2}$ at 1.2~eV. Hoffmann~\textit{et al.}~\cite{Hoffmann:2008} identified this resonance as a $\pi^{*}$ resonance. Our experimental data in this energy region is not reliable; however, theoretical calculations reveal contributions from both linear and bent geometries. A peak in the linear configuration is observed near 0.94 eV (blue curve in Fig.~\ref{fig:s_cs_calc}(a)). The OCS NIR state could also undergoes significant bending, which dominates the dissociation dynamics in this energy range (red curve in Fig.~\ref{fig:s_cs_calc}(a)). The shoulder-like feature observed around 1.4 eV in the S$^{-}$ yield is suggested by our calculations to arise from the bending motion of the transient anion, rather than from the vibrational excitations of the neutral CO fragment. The calculated cross section curves display vibrational structures for the negative ion resonant~(NIR) state in both the linear and bent geometries. We extend our analysis to the subsequent resonances observed in the S$^-$ channel.

The resonance centered at $5$ eV reaches a peak cross section of $1.19\times10^{-19}\text{cm}^{2}$, this value is in agreement with the measurement of Iga \textit{et. al.}~\cite{IGA:1996}. Ab initio calculations, supported by experimental evidence, assign the dominant NIR state to a $\Sigma$ symmetry within the C$_{\infty\text{V}}$ point group (linear geometry)~\cite{kundu_arxiv:2025,Kundu:2024}. For the bent geometry (C$_{\text{S}}$ point group), the corresponding electronic transition occurs from the ground state A$^{\prime}$ to the NIR state A$^{\prime\prime}$. The computed cross sections for dissociation via both the linear and bent NIR states are shown in Fig.~\ref{fig:s_cs_calc}(b). The integrated area under each dissociation curve, derived from both theoretical and experimental cross section data, is summarized in Table~\ref{tab:area_under_the_curve}. In the experimental spectrum, the 5 eV resonance spans the $4-5.6$ eV range, and the area was evaluated across this interval. The experimentally determined area shows excellent agreement with the calculated contribution from the linear NIR state dissociation, whereas the contribution from the bent state is negligible in this energy region. This indicates that the bent NIR state is largely inactive at this resonance. 

The resonance centered at $6.8$ eV exhibits a peak cross section of $2.83 \times 10^{-19}\text{cm}^2$, which is in good agreement with the values found in the literature~\cite{IGA:1996}. Experimental observations with ab initio calculations by Kundu \textit{et al.}~\cite{Kundu:2024} reveal the presence of linear NIR state with $\Sigma$ (dominant) symmetry and bent structure with A$^{\prime\prime}$ symmetry, both contributing to the observed feature around 6.8 eV. In our measured cross section curve, the 6.8~eV resonance spans the $5.6-8.5$~eV electron energy range. The corresponding calculated cross sections reveal four NIR states---two associated with C$_{\infty\text{V}}$ symmetry and two with C$_{\text{S}}$ symmetry (see Fig.~\ref{fig:s_cs_calc}(c)--(d)). We computed the total area under the curve for both linear and bent geometries. A comparison with the experimental data indicates that the NIR states associated with bent geometries contribute significantly, suggesting that these configurations actively participate in the dissociation dynamics at this resonance.

The resonance centered at $10.2$ eV exhibits a peak cross section of $5.43 \times 10^{-20} \text{cm}^{2}$, which is consistent with previously reported values. Not much study on this resonance is available. Our calculations identify a single bent NIR state near 10.1 eV as the primary contributor to this resonance no linear NIR state is available. Our experimental data show a span over the $8.5-13$ eV electron energy range. The agreement between the area under the experimental and calculated cross section curves further supports the assignment. Notably, only the bent NIR state contributes to this resonance. Our calculations indicate that with increasing electron energy, a greater number of bent NIR states become accessible and contribute significantly to the DEA dynamics of OCS.

\subsubsection{\label{subsubsec:s_ipd_dissociation}The IPD process}

Ion-pair production in OCS leading to S$^{-}$ formation typically follows the reaction pathway: $\text{OCS} + e^{-} \rightarrow \text{OCS}^{*} + e^{-} \rightarrow \text{CO}^{+} + \text{S}^{-} + e^{-}$. Here, the incident electron transfers energy to a neutral OCS molecule, promoting it to a superexcited state within the ionization continuum. These superexcited states can be accessed either directly or via coupling with Rydberg states. A minimum threshold energy is required for the incident electron to access the ion-pair states. The ion-pair threshold energy was calculated using the following formula.

\begin{equation*}
    \text{E}_{\text{th}}(\text{CO}^{+}+\text{S}^{-})=\text{D}(\text{\chemfig{OC=S}})+\text{IP}(\text{CO})-\text{EA}(\text{S})
    \label{eq:ip_threshold_s}
\end{equation*}

Here, D denotes the dissociation energy of the \chemfig{C=S} bond, IP is the first ionization potential of CO, and EA represents the electron affinity of sulfur. Using the thermochemical data summarized in Table~\ref{tab:thermochemical_data}, the ion pair dissociation threshold for S$^{-}$ is estimated to be $15.07$~eV. 

To determine the appearance threshold experimentally, the near-threshold region of the S$^{-}$ ion yield curve was fitted using the generalized Wannier threshold law~\cite{Wannier:1953}, which is described by a piecewise function involving a background offset, an amplitude term, the threshold energy, and the Wannier exponent, expressed by the function:

\begin{equation}
    f(\text{E}) = 
    \begin{cases}
        b&\text{ for } \text{E}\leq c\\
        b+a\left(\text{E}-c\right)^{d}&\text{ for } \text{E}\geq c\\
    \end{cases}
\end{equation}

\begin{figure*}
\centering
\includegraphics[scale=0.6]{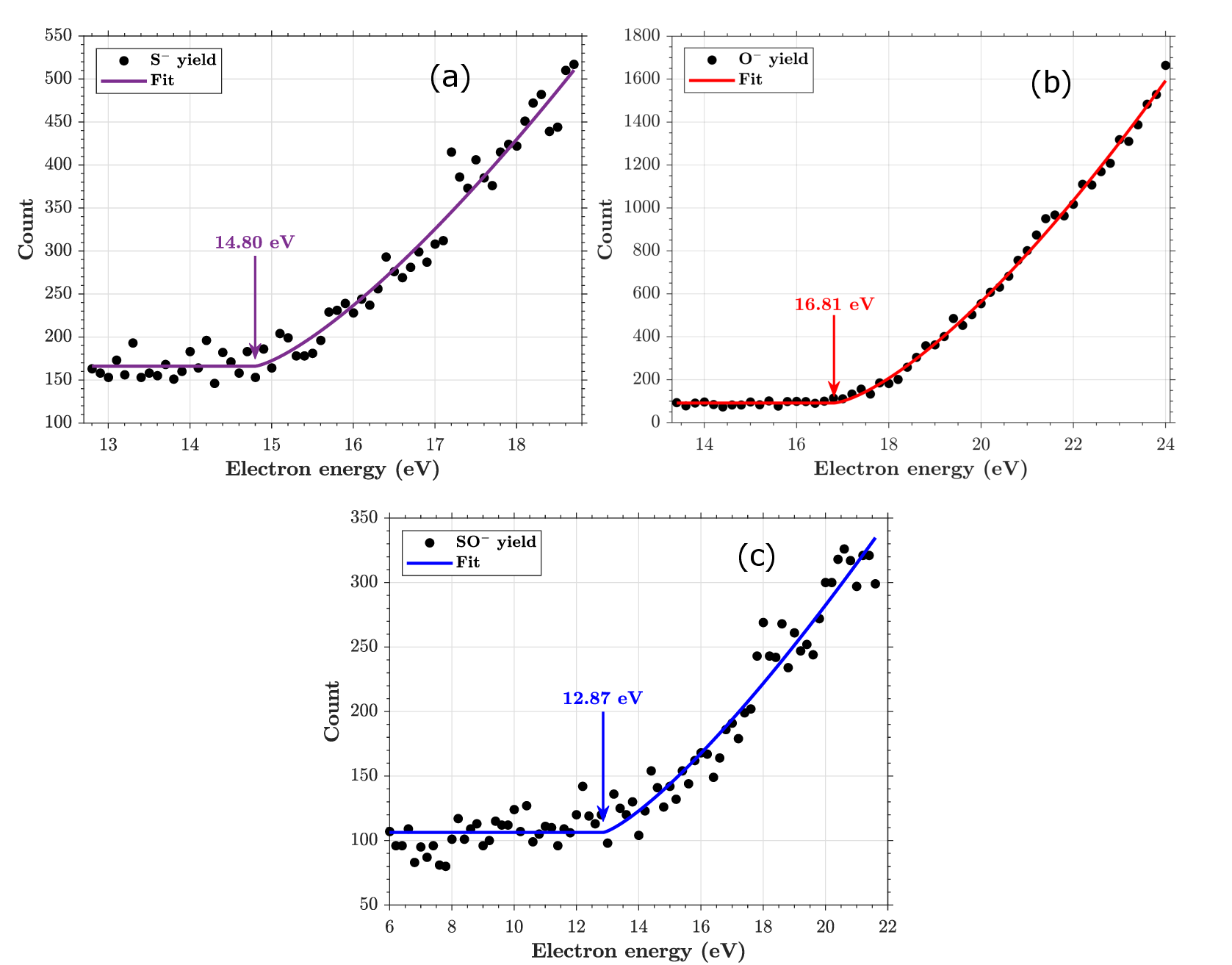}
\caption{\label{fig:ip_thresholds}Ion yield as a function of electron energy in the threshold region for (a)~S$^{-}$, (b)~O$^{-}$ and (c)~SO$^{-}$ ions. The vertical markers indicate the appearance energies obtained from the Wannier law fit.}
\end{figure*}

\noindent In this model, $b$ represents the background offset, $a$ the amplitude scaling factor, $c$ is the ion appearance energy, and $d$ the Wannier exponent. Fitting our S$^{-}$ yield to the Wannier law gives an appearance threshold of $14.80$ eV (see in Fig.~\ref{fig:ip_thresholds}(a)), which is well in agreement with the thermochemical estimate.

Measured cross sections for ion-pair dissociation in OCS are provided in Table~\ref{tab2:cross_sections}. Within this energy range, the cross section remains relatively constant. A representative value for the S$^{-}$ ion production is $8.96 \times 10^{-20}~\text{cm}^2$ at an incident electron energy of $32.3$~eV.

\begin{table}[b]
\caption{\label{tab:thermochemical_data}Thermochemical data for calculation of the Ion-pair threshold}
\begin{ruledtabular}
\begin{tabular}{ll}
Values used in calculation                          & Values (eV)\\
\hline
Bond dissociation energy of \chemfig{SC=O}          & $6.88$~\cite{Liu:2003}\\
Bond dissociation energy of \chemfig{OC=S}          & $3.14$~\cite{Liu:2003}\\
Bond dissociation energy of \chemfig{S-O}           & $7.54$ (CCSD)\\
Electron affinity of C                              & $1.26$~\cite{Myers:1990}\\
Electron affinity of O                              & $1.46$~\cite{Myers:1990}\\
Electron affinity of S                              & $2.08$~\cite{Blondel:2005}\\
Electron affinity of SO radical                     & $1.38$ (EOM-CCSD)\\
First ionization potential of C                     & $11.26$~\cite{Haris:2017}\\
First ionization potential of CO                    & $14.01$e~\cite{Erman:1993}\\ 
First ionization potential of SO radical            & $9.36$ (EOM-CCSD)\\
First ionization potential of CS radical            & $11.55$ (EOM-CCSD)\\ 
\end{tabular}
\end{ruledtabular}
\end{table}

\subsection{\label{subsec:o_so_c_dissociation_channel}O$^{-}$, SO$^{-}$ and C$^{-}$ production}

\begin{figure*}
\centering
\includegraphics[scale=0.6]{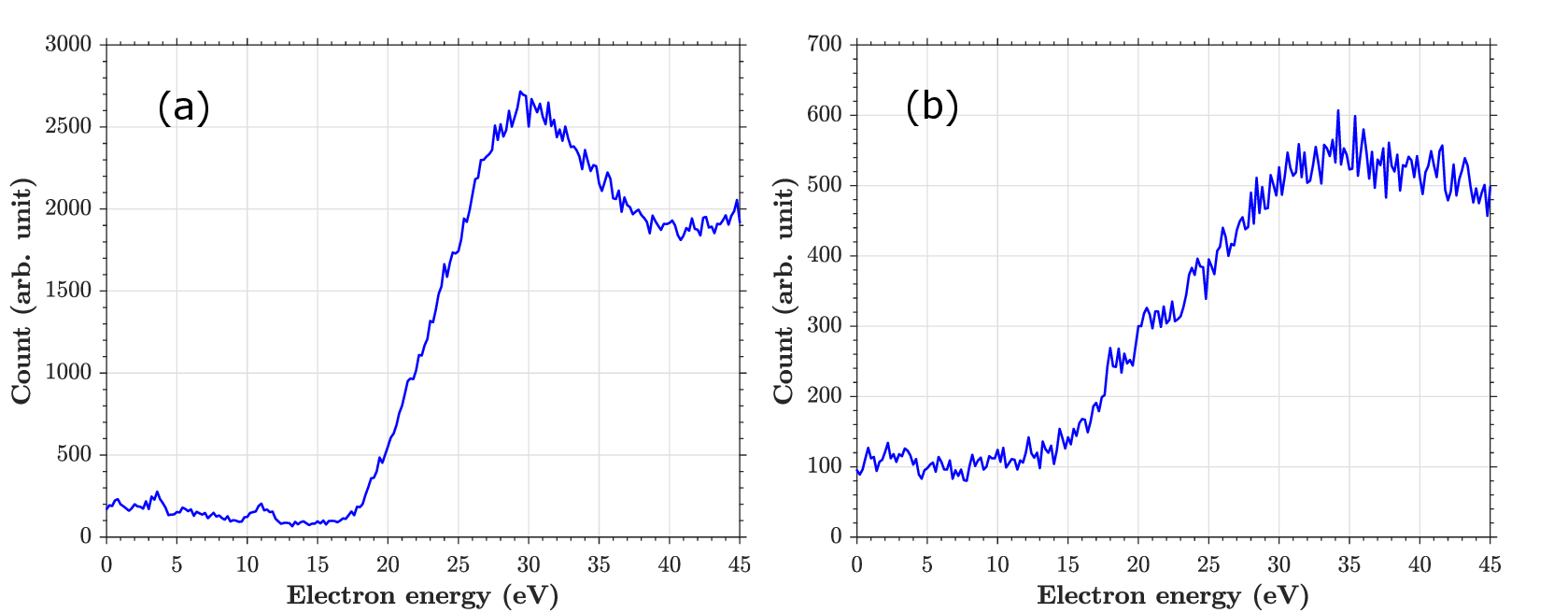}
\caption{\label{fig:o_so_ion_yield}(a)~The ion yield curve for O$^{-}$ ions over the $0-45$~eV incident electron energy range exhibits weak peaks near $3.5$~eV and $11$~eV, indicating possible low-intensity resonances.  
(b)~The ion yield curve for SO$^{-}$ ions over the $0-45$~eV incident electron energy range shows no distinct DEA peaks. Instead, a gradual increase in signal is observed in the ion-pair dissociation region, suggesting that SO$^{-}$ formation occurs primarily through ion pair production.}
\end{figure*}

Hubin-Franskin~\cite{hubin:1976} identified three dominant resonances for the formation of O$^{-}$ near 4.2, 9.6, and 11.0 eV, along with two weaker features at approximately 7.2 and 8.1 eV. Fig.~\ref{fig:o_so_ion_yield}(a) shows our O$^{-}$ ion-yield curve, where only weak features are observed near 3.5 eV and 11 eV in the low-energy region. At higher electron energies, within the ion-pair formation regime, a significant increase in O$^{-}$ ion yield is observed, attributed to the reaction $\text{OCS} + e^{-} \rightarrow \text{OCS}^{*} + e^{-} \rightarrow \text{O}^{-} + \text{CS}^{+} + e^{-}$. The ion-pair threshold for O$^{-}$ formation is evaluated using the following expression: 
\begin{equation*}
\text{E}_{\text{th}}(\text{CS}^{+}+\text{O}^{-})=\text{D}(\text{\chemfig{SC=O}})+\text{IP}(\text{SC})-\text{EA}(\text{O})
\label{eq:ip_threshold_o}
\end{equation*}
Applying the Wannier threshold law to the ion yield data yields an appearance threshold of $16.81$ eV (see Fig.~\ref{fig:ip_thresholds}(b)). The ion pair threshold estimated from thermochemical data using the corresponding relation is $16.97$ eV, which is in good agreement with the value obtained from Wannier’s law.

The SO$^{-}$ fragment is exclusively formed in the ion pair dissociation regime, $\text{OCS}+e^{-}\,\rightarrow\,\text{OCS}^{*}+e^{-}\,\rightarrow\,\text{SO}^{-}+\text{C}^{+}+ e^{-}$. Its formation requires marked bending of the electronically excited, originally linear OCS molecule, facilitating the approach of the O and S atoms to form an \chemfig{S-O} bond, accompanied by the cleavage of both \chemfig{S=C} and \chemfig{C=O} bonds. The corresponding ion yield curve is presented in Fig.~\ref{fig:o_so_ion_yield}(b). The ion pair appearance threshold is 12.87 eV (see Fig.~\ref{fig:ip_thresholds}(c)). The threshold energy for this process incorporates the dissociation energies of the \chemfig{S=C} and \chemfig{C=O} bonds, as well as the formation energy of the \chemfig{S-O} bond. The equation will be as follows:
\begin{eqnarray*}
\text{E}_{\text{th}}(\text{C}^{+}+\text{SO}^{-})=&&\text{D}(\text{\chemfig{SC=O}})+\text{D}(\text{\chemfig{OC=S}})\nonumber\\
&&-\text{D}\left(\text{\chemfig{S-O}}\right)+\text{IP}(\text{C})-\text{EA}(\text{SO})
\label{eq:ip_threshold_so}
\end{eqnarray*}

Using thermochemical data, the ion-pair dissociation threshold for SO$^{-}$ is estimated to be 12.36~eV.

Similarly, C$^{-}$ ions are detected only through the ion-pair dissociation pathway, $\text{OCS}+e^{-}\rightarrow\text{OCS}^{*}+e^{-}\rightarrow\text{C}^{-}+\text{SO}^{+}+ e^{-}$. The ion-pair threshold for the SO$^{+}$ and C$^{-}$ channel is computed using the following relation:
\begin{eqnarray*}
\text{E}_{\text{th}}(\text{SO}^{+}+\text{C}^{-})=&&\text{D}(\text{\chemfig{SC=O}})+\text{D}(\text{\chemfig{OC=S}})\nonumber\\
&&-\text{D}\left(\text{\chemfig{S-O}}\right)+\text{IP}(\text{SO})-\text{EA}(\text{C})
\label{eq:ip_threshold_c}
\end{eqnarray*}
The formation of C$^{-}$ ions is also expected to proceed via a bending motion that facilitates \chemfig{S-O} bond formation before cleavage. Similar observations have previously been reported for carbon dioxide, where the formation of C$^{-}$ ions was observed in the DEA energy range by Wang \textit{et al.}~\cite{Wang:2016}. Due to weak signal intensity, an ion-yield curve for C$^{-}$ could not be obtained. The calculated ion-pair threshold from thermochemical data for this channel is 10.58~eV.

\section{\label{sec:conclusion}Conclusion}

This study presents a comprehensive investigation of DEA and IPD processes in carbonyl sulfide, combining experimental measurements with advanced ab initio theoretical modeling. For S$^{-}$ formation, the DEA dynamics are successfully reproduced by theoretical calculations, with absolute cross sections showing excellent agreement with experiment. The contribution of NIR states—particularly their decay facilitated by coupling between linear and bent geometries via conical intersections. Theoretical analysis indicates that the bending motion plays an increasingly dominant role in governing the dissociation pathways with rising electron energy. However, due to limited electron beam intensity near the 1.2~eV resonance, the specific involvement of linear or bent states in the DEA process at this energy remains inconclusive; nevertheless, the presence of a shoulder in the ion yield suggests the contribution of bent geometry. The resonance near 5~eV is attributed to dissociation through the linear NIR state, while higher-energy resonant features predominantly arise from bent configurations. Consistent with the findings of Kundu~\textit{et al.}~\cite{kundu_arxiv:2025}, our analysis confirms that electron attachment induces significant bending in the OCS molecule and that most dissociation events proceed through these bent geometries.

In contrast, accurate estimation of absolute cross sections for IPD remains significantly more challenging. This difficulty stems from the complex nature of superexcited states and their coupling to the Rydberg continuum, especially in the context of dipole breakdown transitions~\cite{kundu:2023}. Experimental limitations, including weak signals for fragments such as C$^{-}$ and SO$^{-}$, as well as electron beam normalization constraints, further complicate reliable cross section determination. The absolute cross section for S$^{-}$ formation via IPD was estimated by extrapolating from the DEA ion yield, providing an upper bound on contributions up to 45~eV. Theoretically, robust modeling of IPD dynamics necessitates incorporating strong non-adiabatic effects such as Renner–Teller coupling and conical intersections, which significantly influence the fragmentation behavior. These effects require sophisticated multi-surface quantum dynamics approaches and high-accuracy potential energy surfaces with precise electronic coupling elements. The computational complexity becomes particularly demanding when the ion-pair states possess diffuse or Rydberg-like character. Overall, this combined experimental-theoretical approach establishes a valuable framework for understanding electron-induced dissociation processes in polyatomic systems like OCS.

This integrated experimental and theoretical investigation highlights the strong interplay between electronic structure and nuclear dynamics in molecular fragmentation, setting a benchmark for future studies on polyatomic systems undergoing electron attachment. By combining experimental observations with advanced electronic structure and quantum dynamical simulations, this work significantly improves our fundamental understanding of electron-induced processes and their broader importance in chemical reactivity, astrochemistry, and atmospheric chemistry.

\begin{acknowledgments}
SG and NK acknowledge the financial support and experimental facilities provided by IISER Kolkata. 
\end{acknowledgments}

\bibliography{ocs_abs_cs}

\end{document}